\documentclass[%
 reprint,
 amsmath,amssymb,
 prb,
 superscriptaddress
]{revtex4-2}

\usepackage{hyperref}

\usepackage[left=1.8cm, right=1.8cm, top=1.8cm, bottom=2.0cm]{geometry}
\usepackage[english]{babel}
\usepackage[utf8]{inputenc}
\usepackage{amsthm}
\usepackage{mathtools}
\usepackage{physics}
\usepackage{xcolor}
\usepackage{adjustbox}
\usepackage{placeins}
\usepackage[T1]{fontenc}
\usepackage{lipsum}
\usepackage{csquotes}
\usepackage{setspace}

\usepackage{algorithm}
\usepackage[noend]{algpseudocode}
\algblock{Input}{EndInput}
\algnotext{EndInput}
\algblock{Output}{EndOutput}
\algnotext{EndOutput}

\usepackage{siunitx}

\usepackage{multirow}
\usepackage{amssymb}
\usepackage{pifont}

\usepackage{titlesec}
\titlespacing\subsubsection{0pt}{12pt plus 4pt minus 2pt}{0pt plus 2pt minus 2pt}

\begin{document}

\title[Identifying Crystal Structures Beyond Known Prototypes from X-ray Powder Diffraction Spectra]{Identifying Crystal Structures Beyond Known Prototypes\\from X-ray Powder Diffraction Spectra}

\author{Abhijith S. Parackal}
    \affiliation{Department of Physics, Chemistry and Biology, Linköping University, Sweden}

\author{Rhys E. A. Goodall}
    \affiliation{Department of Physics, University of Cambridge, United Kingdom}

\author{Felix A. Faber}
    \email[Correspondence email address: ]{ff350@cam.ac.uk}
    \affiliation{Department of Physics, University of Cambridge, United Kingdom}
    
\author{Rickard Armiento}
    \email[Correspondence email address: ]{rickard.armiento@liu.se}
    \affiliation{Department of Physics, Chemistry and Biology, Linköping University, Sweden}

\date{\today}
\begin{abstract}

The large amount of powder diffraction data for which the corresponding crystal structures have not yet been identified suggests the existence of numerous undiscovered, physically relevant crystal structure prototypes. In this paper, we present a scheme to resolve powder diffraction data into crystal structures with precise atomic coordinates by screening the space of all possible atomic arrangements, i.e., structural prototypes, including those not previously observed, using a pre-trained machine learning (ML) model. This involves (i) enumerating all possible symmetry-confined ways in which a given composition can be accommodated in a given spacegroup, (ii) ranking the element-assigned prototype representations using energies predicted using the Wren ML model [Sci.\ Adv.\ 8, eabn4117 (2022)], (iii) assigning and perturbing atoms along the degree of freedom allowed by the Wyckoff positions to match the experimental diffraction data (iv) validating the thermodynamic stability of the material using density-functional theory (DFT). An advantage of the presented method is that it does not rely on a database of previously observed prototypes and is, therefore capable of finding crystal structures with entirely new symmetric arrangements of atoms. We demonstrate the workflow on unidentified XRD spectra from the ICDD database and identify a number of stable structures, where a majority turns out to be derivable from known prototypes. However, at least two are found not to be part of our prior structural data sets. 

\end{abstract}


\maketitle

\section{Introduction}

Powder diffraction using X-rays (XRD) is a key characterization technique to resolve crystal structures which is used in most materials-related fields.
Refinement based on diffraction data is generally highly successful in yielding precise and accurate structural information \cite{massa2004crystal, cullity1956elements}. However, there are also many challenges associated with this process \cite{cryst7050142, patterson1944ambiguities, harrison1993phase}, leading to the existence of numerous synthesized materials with indexed diffraction peaks for which the precise crystal structures remain unknown \cite{gindhart_power_2018, hybrid}.
The discovery of these unknown structures plays a crucial role in addressing materials design challenges \cite{harris2001contemporary}.

Moreover, the structural information in most theoretical crystal structure databases \cite{jainCommentaryMaterialsProject2013, oqmd, curtarolo_aflowliborg_2012} originate from experimental structures \cite{ICSD, COD}.
Discovering new crystal structures that are not yet included in the existing crystal structure databases is essential for expanding their diversity.
Many machine learning (ML) models for materials rely on these theoretical databases for training data \cite{PhysRevLett.120.145301,chen_universal_2022,goodall_rapid_2022,deng2023_chgnet}.
With a broader selection and structurally diversified set of materials to train on, these algorithms can expedite the discovery of materials with desirable properties. 

A significant volume of literature addresses the structural characterization of XRD patterns.
The first steps usually include separating the contributions from phases present in the sample, locating the Bragg peaks, indexing, and spacegroup classification \cite{hariscsd1996}.
Automated tools have long been in common use for indexing~\cite{visserFullyAutomaticProgram1969, shirley1980data,werner_treor_1985,gilmoreHighthroughputPowderDiffraction2004, altomare2019software, billingeMachineLearningCrystallography2024}.
Many prior works have explored ML models for these and related tasks. LeBras \textit{et al.} used constraint reasoning and kernel clustering for phase identification \cite{lebras_constraint_2011} and Park \textit{et al.}\ developed convolutional neural networks to classify crystal system, extinction group, and spacegroup for a pattern \cite{Park:fc5018}. More recent examples include Bragg peak positioning \cite{liu_braggnn_2022}; crystallographic orientation~\cite{de_oca_zapiain_interpretation_2023}, artifact removal \cite{yanxon_artifact_2023}; novelty detection~\cite{banko_deep_2021}; volumetric segmentation~\cite{sullivan_volumetric_2019}; phase identification and background extraction \cite{bunn_generalized_2015,kusne_high-throughput_2015,ermon_pattern_2015,bunn_semi-supervised_2016,iwasaki_comparison_2017,xue_phase-mapper_2017,xiong_automated_2017,suram_automated_2017,stanev_unsupervised_2018,wang_rapid_2020,lee_deep-learning_2020,lee_data-driven_2021,dong_deep_2021,maffettone_crystallography_2021,szymanskiProbabilisticDeepLearning2021,chen_automating_2021,suzuki_automated_2022,zhdanov_machine_2023,le_deep_2023,fang_application_2023,szymanski_adaptively_2023,drapeau_improving_2023,zhdanov_machine_2023, yue_phase_2024, oppliger_weak_2024}; and a wide range of structural classifications \cite{schuetzke_enhancing_2021,zhao_application_2018,tiong_identification_2020,salgado_automated_2023,chakraborty_deep_2022,chen_crystal_2024,lee_powder_2022,hamza_machine_2023,corriero_crystalmela_2023,zaloga_crystal_2020,suzukiSymmetryPredictionKnowledge2020a,li_composition_2021,vecsei_neural_2019,liu_using_2019,aguiar_crystallographic_2020,suzuki_machine_2018,lee_deep_2023,aguiar_decoding_2019,suzukiSymmetryPredictionKnowledge2020a,corriero_crystalmela_2023}. Hence, there is a rich set of prior works on identifying spacegroup, lattice parameters and compositions from experimental Powder XRD patterns. For further details, see recent reviews covering this topic, e.g., Refs.~\onlinecite{Surdu_review,szymanski_toward_2021,choudhary_recent_2022,hinderhofer_machine_2023}. 

The focus in the present work is on the next step of structural characterization: taking an indexed and classified pattern and derive one or more matching crystal structures with specific coordinates for the atomic positions (see discussions in Refs.~\onlinecite{jones1984crystal,harris2001contemporary, hariscsd1996, brunger1992free,glusker1996crystal,  flack2008use, Altomare:li5027}). The use of ML models to directly solve the inversion problem currently appears to be limited to specific types of systems~\cite{massuyeau_perovskite_2022, oviedoFastInterpretableClassification2019}. The more common approach is to start from reference structures, selected either for compositional similarity or by matching known diffraction patterns, which are then iteratively refined.
The atomic positions within these structures are optimized to align the simulated and experimental XRD patterns, targeting the minimization of the \textit{R}-value \cite{brunger1992free, Ladd-crystallography, Giacovazzo}, a measure of XRD pattern misalignment which is widely recognized to not produce a unique solution. \cite{harrison1993phase}.
It is a common practice to rule out some of the matching structures based on thermodynamic instability, which can be estimated, e.g., by calculations using density-functional theory (DFT) \cite{hohenberg1964inhomogeneous, kohn1965self}.
Numerous techniques, such as particle swarm optimizers, Monte Carlo methods and genetic algorithms, address the challenges of simultaneously optimizing the \textit{R}-value and thermodynamic stability \cite{lonie_xtalopt_2011, ward_three_2015, putz_combined_1999, bush_evolutionary_1995, expo, sheldrick2015shelxt, mcgreevyReverseMonteCarlo2001a}.


The iterative optimization of the positions of $N$ atoms in 3D space to match a powder XRD pattern has a computational complexity of $O(N^3)$. A first step to overcome this combinatorial wall is to restrict the degrees of freedom by leveraging the identified symmetry.
The first-principles-assisted structure solution (FPASS) technique~\cite{hybrid,wardAutomatedCrystalStructure2017} further leverages symmetry information by biasing the assignment to Wyckoff positions~\cite{muller2006remarks} based on occupancies mined from crystal structure databases.
This approach allows a high-throughput workflow to analyze XRD data with minimal manual input.
Griesemer \textit{et al.} \cite{Griesemer} developed a method to use reference structures from the prototypes of known structures within crystal structure databases such as OQMD \cite{oqmd}. A similar approach was used in Ref.~\cite{yangStructureminingScreeningStructure2020}. However, approaches that rely, even just in part, on previously observed structures will have difficulties in identifying structures that belong to completely new prototypes that do not yet exist in any databases \cite{schmidt_large-scale}.

In this work, we present an algorithm to invert XRD patterns into crystal structures with specified atomic coordinates without using data from previously identified structures and prototypes. Our approach uses an efficient systematic enumeration of possible reference structures via a coarse-grained representation of crystal symmetry using Wyckoff positions.
The enumeration of Wyckoff position assignments is restricted based on the composition, spacegroup, and formula units per unit cell, which are available from indexing and classification.
The possible assignments are exhaustively explored up to a spacegroup-dependent limit of unit cell complexity.
The candidate reference structures are then prioritized by a predicted formation energy using the \textit{Wyckoff Representation regressioN} (Wren) ML model that operates on the coarse-grained representation \cite{goodall_rapid_2022}. While there are prior works that have proposed systematic enumeration of structures, e.g., over integer decision variables of coordinate values~\cite{gusevOptimalityGuaranteesCrystal2023a}, our approach can demonstrably reach structures up to relevant complexity since the enumeration over Wyckoff positions aligns well both with the information available from structural analysis in crystallography and the description of crystals in the Wren ML model.

We have implemented the above scheme in a software package \textit{httk-symgen}, which utilizes GPU-accelerated optimizations to refine a cost function related to the $R$-value and interatomic distances on an ensemble of possible candidate matches for a given XRD pattern.
We demonstrate the application of our implementation for inverting XRD to crystal structures on synthetic and experimental patterns, including the identification of two crystal structures on prototypes that are not present in materials databases.

The rest of the paper is organized as follows. In Section II, Background, we define the concepts of Wyckoff prototypes and protostructures, which we consistently use for our coarse-grained description of crystal structures. We also explain how simulated and experimental XRD are matched using the $R$-factor and give an overview of the Wren ML model on which this work is based. Section III, Workflow, describes the steps to invert an XRD pattern. Section IV, Implementation, gives the details on our implementation of the workflow in the \textit{httk-symgen} software. Section V discusses limitations in the applicability of our algorithm and the provided implementation. Sections VI-VIII presents tests on synthetic and experimental XRD. Sections IX and X present our discussion and conclusions.

\section{Background}

\subsection{Wyckoff prototypes and protostructures}

Crystal structures are commonly characterized by the symmetry properties of the lattice and the positions of the atoms.
This leads to the division of all possible crystal structures into 230 spacegroups.
Each spacegroup has a set of Wyckoff sites, labeled using letters in the Latin alphabet (a, b, c, etc.), distinguished based on how the symmetry operators act on them.
The coordinates of a Wyckoff position can be fixed, or there can be one or more freely variable coordinates, allowing a degree of freedom along a line, plane, or in 3D space.
We can obtain the coordinates of the atoms by specifying the Wyckoff position and numerical value for each degree of freedom for the given Wyckoff position.
The Volume of International Tables for Crystallography lists the Wyckoff positions for each spacegroup \cite{Aroyo2011183}.

The concept of ``a crystal structure prototype'' does not have a single, strict definition in literature; it can refer to any representation with a one-to-many relationship to a range of crystal structures with precisely specified atomic coordinates.
In this paper, we use the term \textit{Wyckoff prototype} to refer to the specific kind of prototype that groups crystal structures by spacegroup and a list of occupied Wyckoff labels, i.e., where distinct atoms of unspecified species reside.
Furthermore, we will adopt the name \textit{Wyckoff protostructure} for a Wyckoff prototype with specific elements assigned to all of the occupied Wyckoff sites, but where the precise atomic coordinates allowed according to the degrees of freedom of the occupied Wyckoff sites are not specified.

The Wyckoff prototypes used in this work are based on the formalism established by the AFLOW prototype labels and thus use nearly the same notation \cite{AFLOWLibraryCrystallographic, hicks_aflow-xtalfinder_2020}.
An AFLOW prototype label is a textual representation of a prototype defined as the following series of fields separated by an underscore character (\texttt{\_}): the anonymous chemical formula, the Pearson symbol, the spacegroup number, and the sequence of occupied Wyckoff letters. For example, Calcite structure ($\mathrm{Ca}\mathrm{C}\mathrm{O}_3$) has the AFLOW prototype label \texttt{ABC3\_hR10\_167\_a\_b\_c}. In this work, we specify \emph{Wyckoff prototype IDs} using AFLOW labels, which can be extended into \emph{protostructure IDs} by appending a segment consisting of a colon (\texttt{:}) and a list of elements separated by a dash (\texttt{-}) to assign specific elements to each of the Wyckoff sites. For example, the protostructure ID \texttt{ABC3\_hR10\_167\_a\_b\_e:Ca-C-O} signifies that Calcium atoms occupy the orbits of Wyckoff position \texttt{a}, Carbon atoms the \texttt{b} position, and Oxygen the \texttt{c} position.

We furthermore define that these IDs should be normalized over all possible AFLOW labels that represent the same prototype via transformations of the positions under the coset representatives of the affine normalizer of the spacegroup \cite{bilbao} to use the transformed set whose ID would appear first when sorted on the sum of the alphabetical indices of the Wyckoff letters and, second, lexicographically on the order they appear in the ID. The normalization of our IDs turns them into an origin-independent representation of the respective prototype and protostructure.

\subsection{Matching simulated and experimental XRD}
Structure matching with experimental XRD data is done by identifying a suitable crystal structure and computing a simulated XRD of this reference structure. Then, the value of an \textit{R}-factor is optimized by perturbing the atoms in the reference structure \cite{Ladd-crystallography, Giacovazzo,wardAutomatedCrystalStructure2017}.
The $R$-factor is a unitless measure of a relative mismatch between the peaks of the simulated and experimental patterns, but in the context of refining structures, multiple definitions are in use. To allow direct comparisons with Ref.~\cite{Griesemer} we use their definition:
\begin{equation}
  R = \frac{\sum_{peaks}({I_{exp} - I_{sim}})^2}{\sum_{peaks}{I_{sim}}^2}
\end{equation}
where $I$ is the intensity of the peaks.
Each peak in the simulated pattern is matched to the closest peaks in the experimental pattern as long as they are within \ang{0.15} in $2\theta$.
If multiple peaks are within this range, their intensity is combined before comparison. 

The algorithm for generating simulated XRD patterns from a crystal structure is obtained from De Graef and McHenry \cite{de-graef-2007}.
To summarise, all points in the limiting sphere given by $2/\lambda$ are calculated from the reciprocal lattice of the crystal structure, where $\lambda$ is the wavelength of the X-ray beam.
Then, for every reciprocal point $G_{hkl}$ in this limiting sphere, we calculate $\theta$ using the Braggs condition \cite{eltonXRayDiffractionBragg1966a}.
The atomic scattering factor for a given $s =\sin (\theta) / \lambda$ and an element with atomic number $Z$ is given by:
\begin{equation}
    f(s) = Z - cs^2\sum \limits_{i=1}^N a_i e^{-b_is^2}
\end{equation}
where $a$ and $b$ are the fitting parameters for each species and $c$ is 41.78214. 
The structure factor for a given $hkl$ indices is given by:
\begin{equation}
 F_{hkl} = \sum \limits_{i=1}^N f_i(\frac{\sin\theta_{hkl}}{\lambda})e^{2\pi i (hx_i + ky_i + lz_i) }
\end{equation}
where $N$ is the number of atoms in the unit cell and ($x,y,z$) denotes the atoms' relative (fractional) coordinates. 
The intensity is obtained by multiplying $\abs{F_{hkl}}^2$ with the \textit{Lorentz polarization factor}
\begin{equation}
    I_{hkl} =  \abs{F_{hkl}}^2 \frac{1 +  \cos^2(2 \theta_{hkl})}{ \sin^2( \theta_{hkl}) \cos( \theta_{hkl})}.
    \label{eq: generate_xrd}
\end{equation}

We use an implementation of the algorithm based on the one in Pymatgen \cite{ongPythonMaterialsGenomics2013a} that uses atomic scattering parameters from a table in De Graef and McHenry \cite{de-graef-2007}.

The context largely determines what can be considered a good or bad $R$-value. Comparing $R$-values across different studies is challenging due to differences in the definitions. Previous works that discuss this topic more broadly suggest that a low $R$-value does not necessarily guarantee a correct structure, nor does a high $R$-value disqualify a candidate \cite{harlow_troublesome_1996, spek_what_2018, Griesemer}. Therefore, it is recommended to visually inspect how well the patterns match \cite{smith_quantitative_1987, Toby2006}.
In this work, we primarily use the $R$-value as a guide for quick relative comparisons. Our optimization workflow employs a continuous cost function that is inspired by the $R$-factor (see Sec.~\ref{sec:cost}).

\subsection{Wren}
\label{sec:wren}
As outlined in the introduction, our algorithm uses formation energies estimated using an ML model to guide the choice of reference structures to investigate further. Most available energy-predicting models, such as machine learning interatomic potentials, will not work in the present context since they require knowledge of (at least approximate) atomic coordinates, which are not known at the stage when reference structures are selected.

Hence, an essential component of our work is the Wyckoff representation regression (Wren) ML model \cite{goodall_rapid_2022}.
This model takes as input precisely what we defined above as a Wyckoff protostructure and uses a message-passing neural network to predict the lowest formation energy representation of that protostructure allowed by the degrees of freedom for the occupied Wyckoff positions. An ensemble of models is trained using different random initializations, which increases accuracy and enables an internal representation of the uncertainty of the model as a combination of aleatoric and epistemic uncertainty. 

The present work uses the pre-trained model delivered with Ref.~\cite{goodall_rapid_2022}, which consists of 10 ensemble members trained on the union of the datasets of Materials Project (MP) \cite{jainCommentaryMaterialsProject2013} and Wang, Botti, and Marques (WBM) \cite{wang_predicting_2021}. This combined dataset contains 322,915 materials after cleanup, with 19,312 unique Wyckoff prototypes. The resulting validation error for formation energy prediction was shown to be approximately 30 meV/atom. For more details on the ML model and its performance, see Ref.~\cite{goodall_rapid_2022}.

\section{Workflow}
\begin{figure}
    \includegraphics[width=0.4\textwidth]{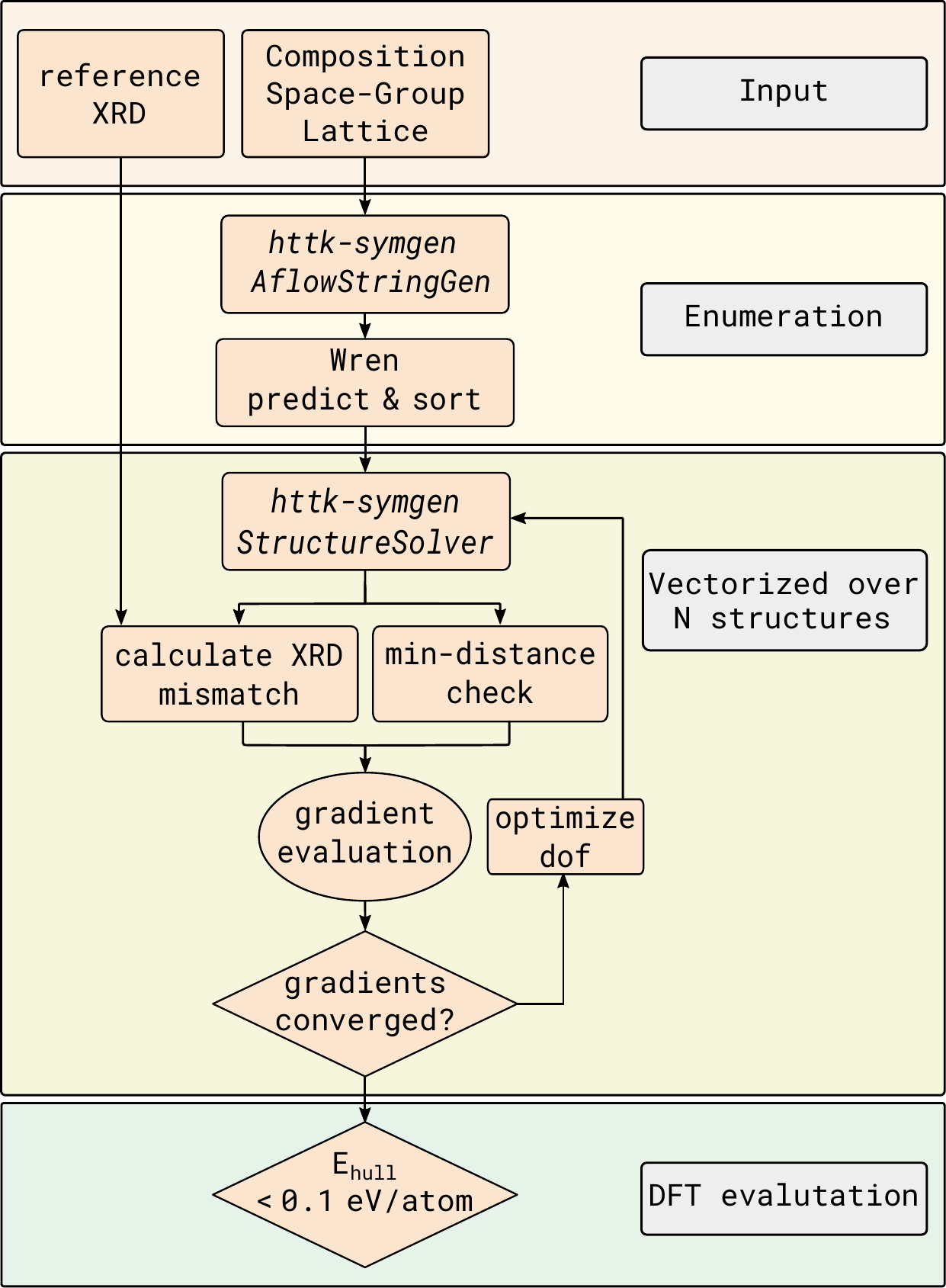}
    \label{fig:workflow}
    \caption{\textbf{Overview of the workflow}: \textit{httk-symgen} contains three major components: \texttt{AflowStringGenerator} method takes composition and spacegroup information and enumerates valid Wyckoff protostructures. This is then ranked and sorted using the Wren model. After sorting, \texttt{StructureSolver} method takes a Wyckoff protostructure and lattice parameters and generates crystal structures by assigning values to the available degrees of freedom.
    Synthetic XRD is generated using this structure and compared with the experimental XRD.
    A structure that matches the experimental XRD and passes the minimal distance checker will be further evaluated using DFT.}
\end{figure}

Our workflow to invert XRD into crystal structures with coordinates takes the compositional and symmetry information (retrieved after indexing) as input. Enumerating the allowed occupations of Wyckoff positions for that spacegroup (see the formulation provided in \ref{sec:enum}), yields a list of Wyckoff protostructures.
These are subsequently ranked and shortlisted based on predicted formation energies using the Wren model.
Each Wyckoff protostructure can be used to construct a crystal structure by populating the degrees of freedom the Wyckoff site allows (detailed in section \ref{sec:Parsing}).
This representation can be passed to an optimization workflow, where the values are filled to minimize a cost function (discussed in \ref{sec:cost}) so that the resultant structure produces a simulated XRD that agrees with the experimental XRD.
An overview of the implementation is provided in Fig \ref{fig:workflow}.

\subsection{Enumeration}
\label{sec:enum}

To enumerate all possible Wyckoff positions that a given composition $\chi_\eta$ can accommodate, we query the multiplicities of each Wyckoff position in a given spacegroup.
Let it be $M = \{m_1, m_2 \ldots m_n\}$ where $m$ is the multiplicity for a given Wyckoff site, and $n$ is the number of Wyckoff positions in that spacegroup.
For a given number of atoms $\eta$,  we reduce the space of $M$ to $M'$ by only taking the sites that match the condition $m_j \leq \eta$ for every $j \in \{1 \ldots n\}$. 

This problem can be formulated as a generalized case of the subset sum problem \cite{gurski_solutions_2020}, where we are to find every subset $S' \subseteq M'$ such that 
\begin{equation}
\label{eq:subset}
    \sum_{i=1}^{n'} km_i = \eta
\end{equation} where $n'$ is the size of $M'$, and $k$ is zero or some positive number.

For a set of elements in $M'$, numerous solutions can satisfy Eq(\ref{eq:subset}).
Each atom type in the composition allows independent solutions.
When chained together, ensuring no two atoms fill a Wyckoff site with zero degrees of freedom (detailed in section \ref{sec:generate}), these solutions lead to a series of Wyckoff protostructures.
The enumeration process can be computationally demanding, especially for compositions with numerous atoms, and its complexity majorly depends on the selected spacegroup.
This problem can be framed as the non-negative integer solutions of the linear diophantine equation \cite{diophantine}.

\subsection{Cost function}
\label{sec:cost}

We use the \textit{R}-value to report how well our simulated patterns match the experimental ones.
However, the optimization part of the workflow requires a more continuous and well-behaved cost function to be stable.
Therefore we use an alternative cost function that measures the overlap between two XRD profiles by mapping a pseudo-Voigt profile over the experimental and simulated peaks (denoted as $\mathbf{f}$ and $\mathbf{g}$, respectively) so that peaks are represented as a continuous vector and calculating
\begin{equation}
    C_{\mathrm{xrd}} = 1 - \frac{\mathbf{f} \cdot \mathbf{g}}{\left \| \mathbf{f} \right \|_2 \left \| \mathbf{g} \right \|_2}.
    \label{eqn:Xrd_eq}
\end{equation}
One of the advantages of this cost function is that it is bounded between $(0,1)$, allowing us to use a fixed threshold to assess the similarity.

We also define another objective function, which ensures that the distance between atoms is not closer than a defined threshold.
We use the average of the sum of Wigner-Seitz radii \cite{wignerredii} of two atoms as this threshold, denoted as $D_{\text{min}}$.
The minimum distance cost function can then be defined as
\begin{equation}
    C_{\mathrm{distance}} = \sum_{i=1}^{n} \sum_{j=i+1}^{n} \max(0, D_{(i,j)}^{\text{min}} - D_{(i,j)})/2
\label{eqn:distance_eq}
\end{equation}
where $D_{i,j}$ is the distance between two atoms $i$ and $j$ in the crystal structure.

In our implementation, a gradient descent optimization is used to identify the minima of the combined XRD and distance cost functions.
The minima are reached when the gradients with respect to the degrees of freedom converge, mirroring the convergence to a local structural minimum.
This search is executed multiple times with varied initializations, and the outcomes are contrasted in order to identify the global minima.
More details are provided in section \ref{sec:xrdopt}.

\section{Implementation}

We have implemented the workflow described in Section III in a subpackage \textit{httk-symgen} of the High-Throughput Toolkit (\textit{httk}) software package \cite{armientoDatabaseDrivenHighThroughputCalculations2020a}.
This software package has two significant components. \texttt{AflowStringGenerator} handles the enumeration and serialization of Wyckoff positions, and \texttt{StructureSolver} handles operations on a given  Wyckoff protostructure.

\subsection{Generating candidate protostructures}
\label{sec:generate}

\texttt{AflowStringGenerator} takes chemical composition and spacegroup and returns a list of Wyckoff protostructures.
This decoration is constructed by formulating the combinatorial search as a modified subset sum problem.
A recursive search function is implemented, which iterates over the multiplicities of Wyckoff sites for a given spacegroup and checks if any combination of summed multiplicity values would be equal to the desired composition (number of atoms). 
More details on the algorithm for the enumeration are given in the supplementary materials, Ref. \onlinecite{supplementary}.

As an example, consider the case where we have to find the number of ways to accommodate 8 atoms in spacegroup 61. Spacegroup 61 can accommodate three Wyckoff positions, denoted by the Wyckoff letters $a$, $b$, and $c$.
Site $a$ and $b$ have 4, and $c$ has 8 multiplicites respectively.
There are two different ways to put eight atoms from this spacegroup, '$ab$' [4+4] and '$c$' [8]. 
One has to be careful of the added constraint that if a Wyckoff site has no degrees of freedom, it cannot be reused again in the solution. For example, we cannot take '$aa$' because the Wyckoff site $a$ does not have any degrees of freedom. Therefore, trying to put 8 atoms in two $a$ sites would mean that more than one atom would sit on exactly the same coordinates.

The \textit{httk-symgen} implementation includes lookup information of symmetry operations, Wyckoff letters and multiplicities, and affine transformations for the origin choices for the standard settings as listed in Volume A of the International Tables for Crystallography \cite{ Aroyo2011183}.
Our lookup tables are based on the data available via python library \texttt{PyXtal} \cite{pyxtal}, which we have modified slightly with the help of the \textit{Bilbao Crystallographic Server} \cite{bilbao} to match the information in the International Tables.

The generated assigned Wyckoff prototype strings are passed through the pre-trained Wren model. As explained in section \ref{sec:wren}, the pre-trained Wren ensemble comprises ten models trained to converge from different random initializations.
A standard deviation is calculated from the ensemble predictions, which is then subtracted from the mean predicted formation energy to yield an uncertainty-adjusted value.
This adjusted value sets a baseline for the formation energy of the protostructures and is used to shortlist the enumerated protostructures.

For efficient evaluation, protostructures with uncertainty-adjusted predicted formation energy exceeding 40 meV/atom above the lowest prediction are excluded.
This threshold, while user-adjustable, is chosen based on the mean absolute error of Wren.
Wren demonstrates better accuracy the closer the prediction is to the convex hull of thermodynamical stability in a sufficiently sampled composition \cite{goodall_rapid_2022}. 
This is a particularly useful property in the context of XRD inversion, where candidates are predominantly located in proximity to the convex hull.

\subsection{Parsing assigned Wyckoff protostructures}
\label{sec:Parsing}
A core feature of \textit{httk-symgen} is the ability to realize a Wyckoff protostructure and create an internal representation that returns a filled cell from a list of values that fills the degrees of freedom allowed by each Wyckoff position, provided by the method \texttt{StructureSolver}.
As an example, consider the protostructure ID \texttt{ABC3\_oP40\_61\_c\_ab\_3c:As-I-O}.

Referring to the previous example, spacegroup 61 accommodates three Wyckoff positions: $4a$, $4b$, and $8c$, with the number indicating multiplicity.
$4a$ and $4b$ do not have any degrees of freedom, while $8c$ is a general position with three degrees of freedom.
Referring to the protostructure ID, the inference to be made here is that 8 Arsenic atoms occupy Wyckoff site $c$, 8 Iodine atoms split between sites $a$ and $b$ (4 atoms in site $a$ + 4 atoms in site $b$), and 24 Oxygen atoms are distributed across three $c$ sites (8+8+8), summing up to 40 atoms in the unit cell.
Therefore, one would have to specify (3)+(0+0)+(3+3+3) or 12 numbers between $[0,1]$ to define the positions of every atom in this structure uniquely.

Every atomic arrangement permitted by this prototype can be explored by adjusting these 12 values.
The desired structure is determined by using this list as input for an optimizer and minimizing specific objective functions, which in this case are the XRD mismatch and distance cost function.

\subsection{Structure refinement using XRD data}
\label{sec:xrdopt}

Our starting point for XRD inversion is an experimental XRD pattern for which the unit cell parameters, the reduced composition, scaled intensity, and $2\theta$ values of the diffraction peaks are known.
This is the information commonly available in databases of otherwise unidentified patterns. 

After obtaining these unit cell parameters and symmetry information, we pass this to \texttt{AflowStringGenerator}.
This yields a list of Wyckoff protostructure IDs, which are passed through Wren for energy prediction. 
The sorted list of Wyckoff protostructure labels is then truncated using the cutoff criteria, and the shortlisted candidates are parsed using the \texttt{StructureSolver} method, which identifies the degrees of freedom for each Wyckoff site in the candidate prototype. 
For each candidate prototype, $n$ different initial structures are generated, using \textit{Latin-hypercube} \cite{loh1996latin} spacing implemented in \texttt{Scipy} \cite{2020SciPy-NMeth} for optimal sampling of the search space.

The cost functions described in equations \ref{eqn:Xrd_eq} and \ref{eqn:distance_eq}, as well as the code for generating simulated XRD,
Eq (\ref{eq: generate_xrd}) is written in the numerical analysis library, \texttt{Jax} \cite{jax2018github, deepmind2020jax}.
Since the problem is framed and written in an ML library that targets parallel computation, we can evaluate these $n$ random structures in parallel on a GPU.

This capability facilitates the simultaneous identification of the global minima for cost functions $C_{\mathrm{distance}}$ and  $C_{\mathrm{xrd}}$ simultaneously over a grid $(n,dof)$, where $(dof)$ denotes the degrees of freedom associated with the Wyckoff prototype.
We employed a gradient descent methodology \cite{gradient-descent} for the optimization of atomic positions in each structure, leveraging the Adam optimizer \cite{kingmaAdamMethodStochastic2017} available in \texttt{Optax}\cite{deepmind2020jax} with a set learning rate of 0.001. 

The gradient convergence, as detailed in the Workflow section \ref{sec:cost}, serves as our termination criterion.
Post-convergence, we filter for unique atomic configurations and rank them by their \textit{R} values before proceeding with DFT calculations for formation energy.

We use the electronic structure code VASP (version 5.4) \cite{KRESSE199615, kresseEfficientIterativeSchemes1996a} for DFT calculations, using the INCAR settings and pseudo potentials chosen to comply with Materials Project \cite{jainCommentaryMaterialsProject2013} settings.
Projector augmented wave pseudo-potentials \cite{kresseUltrasoftPseudopotentialsProjector1999, blochlProjectorAugmentedwaveMethod1994} were used with the Perdew-Burke-Ernzerhof generalized gradient approximation (PBE/GGA) functional (\cite{perdewGeneralizedGradientApproximation1996}). 
The \texttt{MaterialsProject2020Compatibility} \cite{jainFormationEnthalpiesMixing2011a} correction scheme implemented in Pymatgen \cite{ongPythonMaterialsGenomics2013a} was applied to the final DFT energy. We used the High-Throughput Toolkit (\textit{httk}) \cite{armientoDatabaseDrivenHighThroughputCalculations2020a} to manage the calculations.

\begin{figure}
  \includegraphics[width=\linewidth]{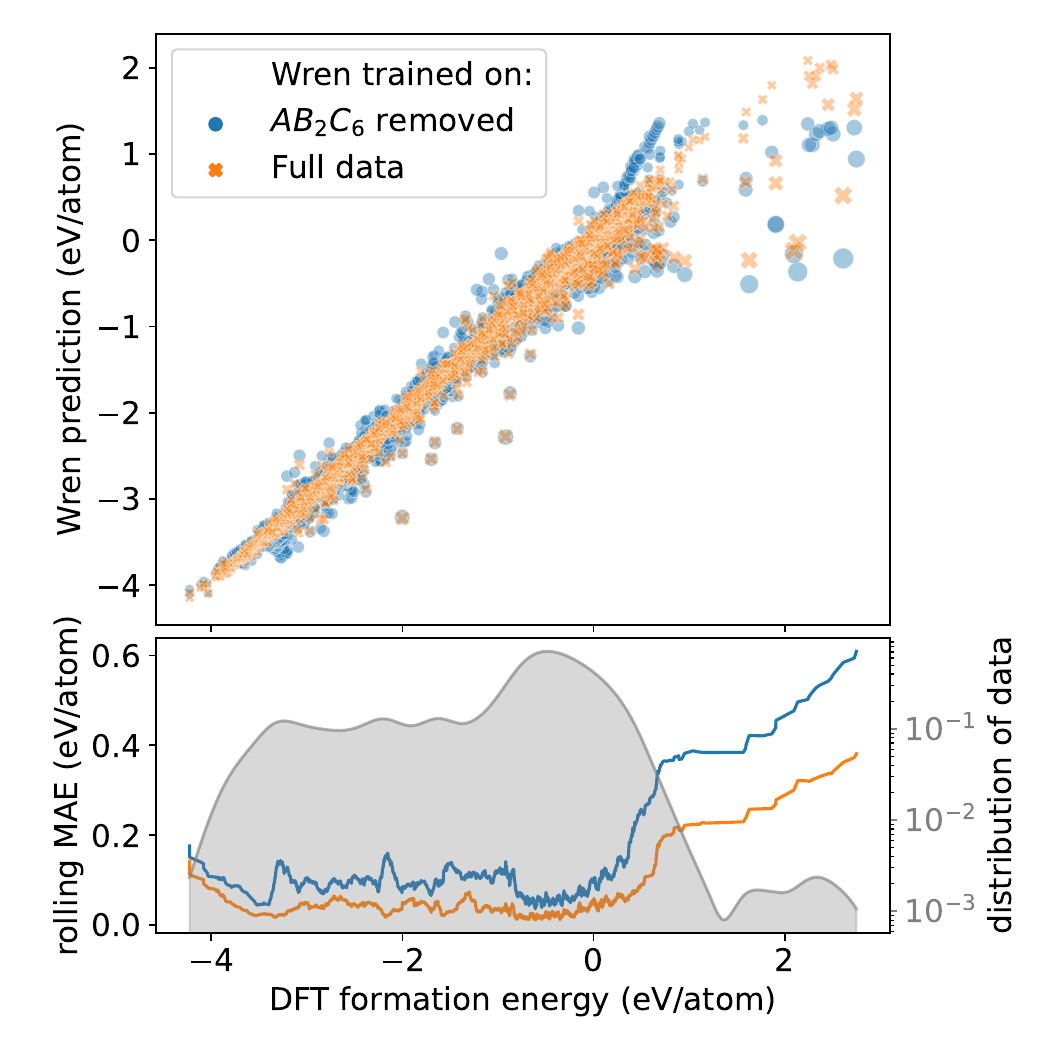}
  \caption{(Upper) A scatter plot comparing the predictive accuracy of Wren models trained on two different datasets: one on the complete WBM+MP dataset and the other omitting all data points with the composition $AB_2C_6$.
  The size of each point represents the level of predictive accuracy, with larger points indicating higher errors.
  (Lower) The rolling MAE of the two models as a function of the DFT formation energy and, in the background, a grey kernel density estimate (KDE) plot of the data distribution. 
  The model trained on the complete dataset has an error of 37 meV/atom, while the model trained on the dataset with the excluded composition reports an error of 88 meV/atom. Wren displays higher accuracy for protostructures with lower formation energy, and its accuracy generally goes down as the formation energy increases.}
  \label{fig:odp}
\end{figure}

\section{Limitations and challanges\label{sec:limitations}}

In this section, we clarify the role of this work as a component of frameworks for large-scale automated structural characterization using XRD by discussing some of the limitations and challenges of applying our work in that context. This kind of use has recently seen increased interest as part of the transition into highly automatized synthesis and characterization, see, e.g., Ref.~ \cite{szymanski_autonomous_2023}.

\paragraph{Uniqueness of solution.} The algorithm produces a prioritized list of solutions that all match a provided XRD, ordered from the lowest predicted formation energy up to a cutoff value related to the accuracy of the energy predictions. In some cases, the result is a single solution, but there may be more, especially if the experimental data is of a nature that does not allow the requirements on the $R$-value match to be very stringent. In such cases, reducing the results into a single true solution requires further in-depth theoretical or experimental analysis, which may not be easy to automate and can require significant effort. Strictly speaking, the same kind of deeper characterization is required also to confirm that the true solution has been found even if there is a single well-fitting identification when there are limits on how exhaustively all possibilities have been screened (e.g., see discussion on complexity below). Nevertheless, the XRD pattern inversion problem has long been known not to be uniquely solvable, and our algorithm arguably handles the ambiguity as well as is possible under the circumstances by finding all solutions matching the provided information. This differs from other tools that are deliberately designed to be biased towards those solutions that share features of previously identified structures.

\paragraph{Bounds on structural complexity.} Our systematic enumeration of protostructures enables an efficient coarse-grained screening of candidates. However, the complexity of the brute force exploration of Wykoff positions in a given spacegroup is similar to the subset sum problem, which is identified to be an NP-hard problem. The combinatorics are especially challenging for compositions with many atoms in spacegroups with a large number of Wyckoff positions with at least one degree of freedom (e.g., 47 and 123). For example this class includes many molecular crystals, which end up having many atoms assigned to general Wyckoff positions in spacegroups representing fairly low symmetry. Hence, in practice, the enumeration has to stop at some level of complexity set by the number of occupied Wyckoff positions and the number of atoms in the unit cell. In this work we have set those limits at 16 assigned Wyckoff positions and 64 atoms per unit cell, which are chosen based on that structures beyond these are not well represented in the Wren training data \cite{goodall_rapid_2022}, and we therefore suggest not going beyond these limits in \textit{httk-symgen}. These bounds preclude finding structures of arbitrary complexity; however, the user can choose the limits based on available computational resources. 

\paragraph{Need for initial indexing, spacegroup identification} The method as presented relies on a prior step of accurate indexing and spacegroup identification. A reasonably small deviation in lattice shape significantly affects the cost function defined in Eq.~(\ref{eqn:Xrd_eq}). Also, XRD patterns may have been obtained under high pressure, which affects subsequent analysis of candidates, e.g., by DFT relaxations using zero external stress.

\paragraph{Presence of multiple phases, disorder, and background contributions.} Our present implementation assumes the provided powder XRD is for a single ordered phase with no significant background contribution, possibly by first taking it through some form of preprocessing. As discussed in the introduction, much prior literature (many involving ML models) deals with the complex problem of separating contributions from multiple phases. However, this problem has been argued to remain a major challenge for automated XRD analysis \cite{PRXEnergy.3.011002}. There is nothing inherent to our algorithm preventing a future extension to consider possible mixtures of enumerated reference structures, and such an extension could help further address these challenges. However, such an extension would significantly increase the computational effort and has not yet been implemented.

\section{Testing generalizability of energy predictions}
Predictive ML models are known to be accurate primarily for data similar to the training dataset, which often is discussed in terms of interpolative vs.\ extrapolative use of models.
In this section, we investigate the out-of-dataset predictive power of Wren in the context of this work by creating two new retrained versions.

The first model is trained only on a subset of the original dataset where we have removed an entire anonymized composition $AB_2C_6$ from the original dataset, which results in 314,554 materials. A technicality of this test is that it would be expensive to train the model to the same level of convergence as the original pre-trained Wren model provided with \cite{goodall_rapid_2022}, given that it will be used only for this test. Hence, we run the training only for 50 epochs, where we observe that when used with 20 ensembles, the overall performance is on par with Wren. We then create a second model trained on the entire dataset used for the original Wren in the same way as the first model (50 epochs, 20 ensembles).

We can now compare these two models for protostructures with the composition removed from the training of the first model, $AB_2C_6$. In this domain, the first model error comes out to 88 meV/atom (see Fig.~\ref{fig:odp}), whereas the second model gives an error similar to the original Wren, 37 meV/atom. The conclusion is that the restricted model can generalize well into the space of an entire composition that is not part of its training data without uncontrollably large errors. Rather, the error for out-of-dataset predictions is thus still below the reported MAE of DFT calculations on general chemistries and structures (ca 100 meV/atom \cite{kirklin_open_2015, aykolThermodynamicLimitSynthesis2018a}). Hence, the remaining accuracy is sufficient for the enumeration-based screening workflow of this work.

\section{Tests of synthetic XRD inversion}

To evaluate the performance of the complete workflow in identifying the atomic positions belonging to a previously unseen prototype from an XRD pattern, we randomly chose a few materials from the materials project dataset. These materials were part of the $AB_2C_6$ space that is excluded from the training data in the retrained models discussed in the previous section.
We chose structures that are evaluated to be on the convex hull of thermodynamical stability.
Then, we simulated the XRD patterns for these structures.
We were able to reproduce the original crystal structure from which the XRD was generated by starting only from the simulated XRD and the composition.
The two examples are discussed in more detail below, with a summary presented in Table \ref{table:1} (more data is available in the supplementary materials, Ref.~\onlinecite{supplementary}).

In the first test, we start from a synthetic XRD for {\textbf{$\mathrm{Li}_2\mathrm{MnF}_6$} (mp-$752936$)} and the knowledge that it is in spacegroup 150. Using \texttt{AflowStringGenerator}, we find that this composition and spacegroup allow 7150 different enumerated protostructures. 
Ranking the Wyckoff protostructure IDs using Wren and sorting them based on the aleatoric and epistemic uncertainty (explained in workflow) and using a cutoff of 0.04 eV/atom from the lowest predicted energy gives a shortlisted number of 180 candidates.
The original protostructure ID is \texttt{A6B2C\_hP27\_150\_3g\_ef\_ad:F-Li-Mn}, and it was predicted as the most likely structure (at the top of the sorted list) in the Wren ranking. Hence, the number of protostructures to take to the next step of matching the $R$-value using \texttt{StructureSolver} is brought down by the screening with Wren from 7150 to 180, i.e., in this case, the screening of formation energies reduces the computational effort of the following step with a factor of 40.

The remaining list of protostructures, sorted by the energy predicted by Wren, is sequentially processed using \texttt{StructureSolver}. We initialize 512 random structures for a given protostructure and optimize them to match the XRD profile. For each protostructure, multiple instantiations among the 512 structures relax into the same structure. Duplicate structures are trivially excluded by comparing the values of the degrees of freedom and picking the structure corresponding to the best \textit{R}-value. For this particular case, the \texttt{StructureSolver} screening results in three protostructures \texttt{3g\_ef\_ad}, \texttt{3g\_2e\_ad} and \texttt{3g\_2e\_bd} reproduced the original XRD. It may be sufficient to use DFT to discern the most stable of the structures.

In the second test, we use a synthetic XRD for {$\mathrm{Sb}_2\mathrm{SrO}_6$ (mp-$9126$)} and the knowledge that it is in spacegroup 162. Enumeration gives in this case that the composition and spacegroup only allow 42 protostructures.
Wren is used to predict formation energies, and the cutoff criterion is applied to yield nine shortlisted Wyckoff protostructures.
The original structure has a protostructure ID \texttt{A6B2C\_hP9\_162\_k\_c\_b:O-Sb-Sr} with two degrees of freedom appeared as the third-ranked option. 
In this case, solving the structure requires optimizing only two degrees of freedom, which exactly reproduces the synthetic XRD. The next best structure has an $R$-value greater than 0.15.

Our tests have been run on an NVIDIA A40 GPU, and the Wren formation energy predictions are found to generally, on average, take approximately 0.002 seconds per protostructure. The processing with \texttt{StructureSolver} takes around 30 seconds per protostructure. The computational load can be scaled up more or less perfectly in parallel onto multiple GPUs. This is helpful for potentially solving a large number of XRD peaks that are indexed to sufficient accuracy.
Analysis of an additional 25 structures is provided in the supplementary materials, Ref.~\onlinecite{supplementary}. 

\begin{table}
\begin{tabular}{llllrrrrrr}
     mp-id & composition & spgp & enum & cutoff & rank \\ \hline
 mp-752936 & $\mathrm{Li}_2\mathrm{MnF}_6$ & 150 & 7150 & 180 & 1  \\
 mp-9126 & $\mathrm{Sb}_2\mathrm{SrO}_6$ & 162 & 41 & 9 & 3  \\
\end{tabular}
\caption{Summary of the highlighted structures. 'enum' represents the total number of enumerated candidates, 'cutoff' is the number of candidates left after applying a cutoff of 40 meV above the lowest prediction. 'rank' shows at what position the structure is ranked based on predicted formation energy.}
\label{table:1}
\end{table}

\begin{figure*}
    \centering
    \includegraphics[width=0.8\textwidth]{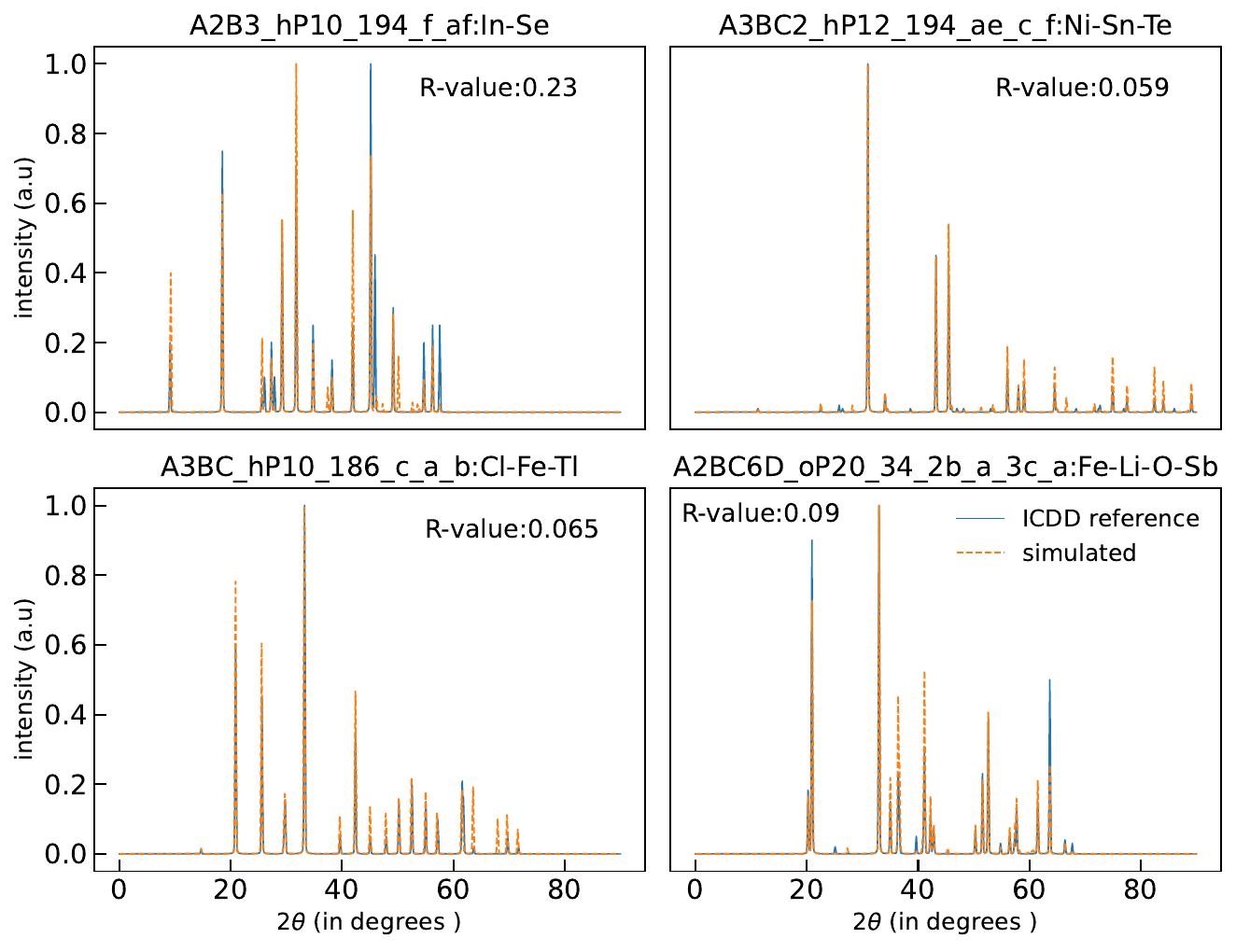}
    \caption{\textbf{Experimental XRD resolved using \textit{httk-symgen}}. Four structures are shown to be solved using our workflow. The blue line shows the experimentally obtained peaks fitted over a pseudo-voigt function. Yellow-dashed lines show the XRD profile generated using structures solved \textit{httk-symgen}. \textit{R}-value is calculated only in the region where ICDD data is available; regions outside are removed from the plot.}
    \label{fig:experimental_results}

\end{figure*}

\section{Tests of experimental XRD inversion}

\begin{table*}[htbp]
\makebox[\textwidth][c]{
\begin{tabular}{ |p{2cm}||p{2cm} |p{7cm}|p{1cm}|p{2cm}|p{2cm}|}
 \hline
 ICDD-ID & composition & protostructure ID & known & DFT hull dist (meV/atom) & R-value  \\ \hline
 
     00-044-1073 & $\mathrm{LiSbFe}_2\mathrm{O}_6$ & A2BC6D\_oP20\_34\_2b\_a\_3c\_a:Fe-Li-O-Sb & no & 9 & 0.090 \\
     00-024-1293 & $\mathrm{TiFeCl}_3$  & A3BC\_hP10\_186\_c\_a\_b:Cl-Fe-Tl & yes & -5 & 0.065 \\
     00-060-0238 & $\mathrm{SnTe}_2\mathrm{Ni}_3$ &  A3BC2\_hP12\_194\_ae\_c\_f:Ni-Sn-Te & no & 7 & 0.059 \\
     00-040-1408 &$\mathrm{In}_2\mathrm{Se}_3$ & A2B3\_hP10\_194\_f\_af:In-Se & yes & 55 & 0.230 \\

  \hline
\end{tabular}
}
\caption{Summary of the crystal structures identified from experimental XRD. The 'DFT hull dist' in the table is the positive or negative distance to the known convex hull of thermodynamic stability from Materials Project; i.e., when negative, it indicates a structure that is stable with respect to the competing phases, and thus that the known hull needs to be updated with the listed structure. The 'known' column means that the prototype is present in at least one of the databases used to train Wren, i.e., MP or WBM.} 
\label{table:2}
\end{table*}

In this section, we use the proposed framework to identify crystal structures from experimental XRD. We first validate the framework by reproducing already identified structures in the RRUFF project database \cite{LafuenteDownsYangStone+2016+1+30} and then apply the same methods on previously unidentified powder diffraction data in the ICDD PDF-2 2012 database \cite{gates-rector_blanton_2019, ICDDpdf2}. The closed nature of the latter imposes technical limitations on large-scale access to the atomic coordinates of identified structures. Hence, we were not able to devise a test of the ability of our framework to recover already identified structures in this database. This highlights the important contribution of open datasets, such as the one provided by RRUFF.


\begin{figure*}
    \includegraphics[width = 0.7\textwidth]{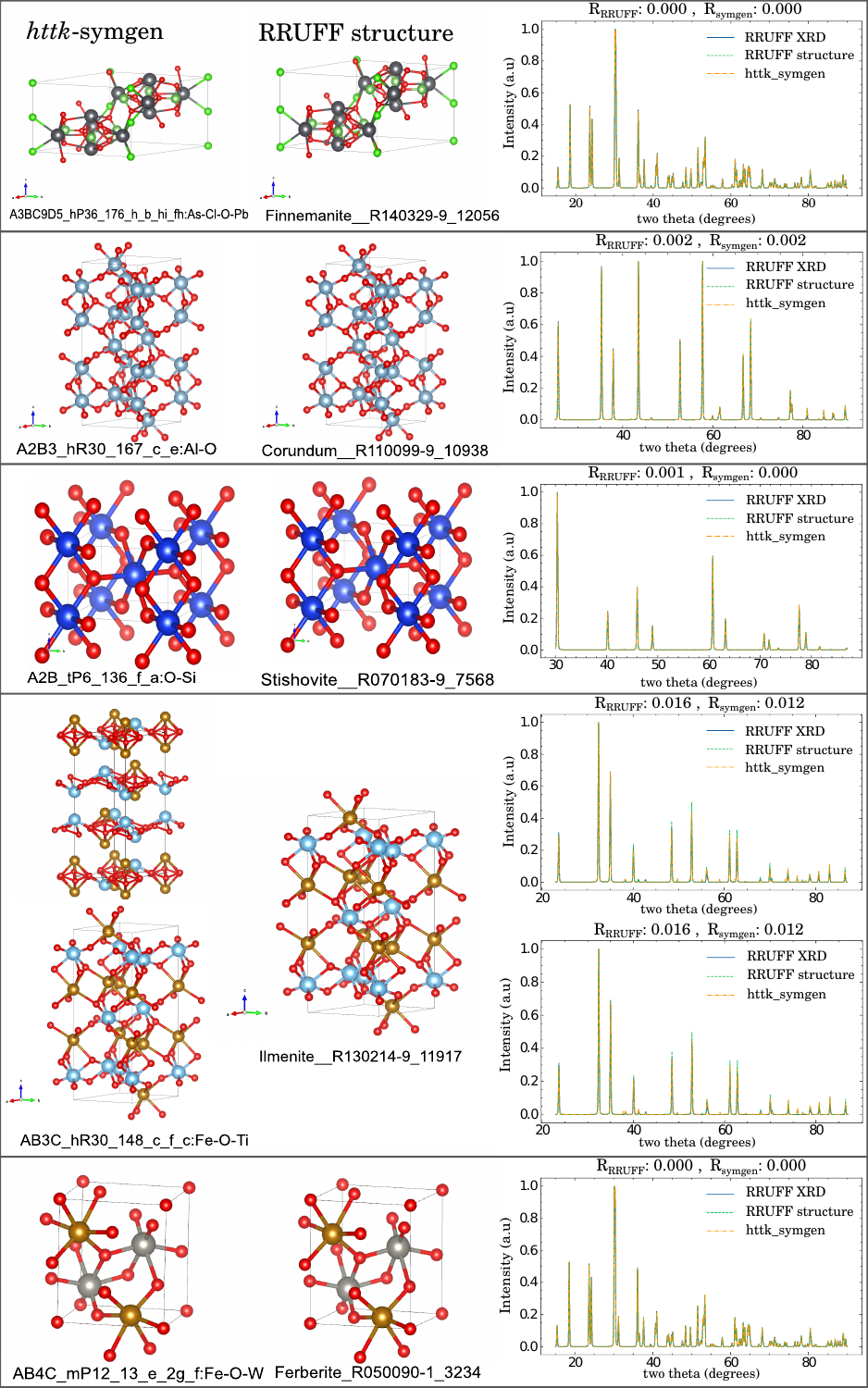}
    \caption{A few examples of structures identified from experimental XRD patterns in the RRUFF dataset.
    Each row corresponds to an entry in RRUFF dataset, showing (first column) all the candidate structures identified by our algorithm; (second column) the crystal structure in the RRUFF dataset, regarded here as the true one; (third column) comparison of the synthetic XRD from our solutions and the experimental XRD in RRUFF. For the fourth row, our algorithm found two solutions with the same good R-value of 0.012. For the fifth row, we found 5 solutions and only show the one with the lowest R-value here (the others are included in the supplementary materials \cite{supplementary}).}
    \label{fig:rruff}
\end{figure*}

\subsection{Test of Recovery of structures in RRUFF}
\label{sec:rruff_para}

We start from the RRUFF project database retrieved on Jan 24, 2024, with 2902 XRD patterns. The dataset is filtered to remove disordered structures (1992), structures that are difficult to automatically process (e.g., internal inconsistencies, unusual symmetry settings, or where the RRUFF data indicates low confidence; 447) and structures of complexity beyond what our present implementation can handle (163) (see Sec.~\ref{sec:limitations}). This leaves 300 entries. We used DFT calculations using the PBE functional \cite{perdewGeneralizedGradientApproximation1996} to relax all these structures and kept only the ones we could confirm to have reasonable stability at zero temperature (i.e., less than 100 meV/atom above the Materials Project \cite{jainCommentaryMaterialsProject2013} convex hull of thermodynamical stability). The end result is a data set of 189 structures suitable for testing our algorithm. (More details on the filtering are available in the supplementary materials, Ref.~\onlinecite{supplementary}).

Our implementation enumerates and identifies crystal structure candidates for the remaining 189 XRD patterns. We count the inversion as a success if the known structure in RRUFF appears anywhere in our final remaining candidates after applying the energy screening cutoff of 40 meV/atom above the energy of the structure with the lowest predicted energy and has an $R$-value match better than a chosen cutoff of less than 0.1. The result is that 186 out of the 189 structures are recovered. Figure \ref{fig:rruff} shows some examples of these recovered structures. However, we stress that this test only demonstrates the ability of our algorithm to reproduce structures from experimental XRD that are similar to those available in the training data set (which is very likely to hold for all the RRUFF structures). Also, the workflow yields multiple candidate structures for a given XRD pattern, (e.g. Ilmenite\_R130214-9\_11917 in Fig \ref{fig:rruff}), which required additional DFT evaluations. More details on this test are available in the supplementary materials, Ref.~\onlinecite{supplementary}.

\subsection{Recovery of unidentified ICDD structures}
\label{sec:unidentified_ICDD}
In this section, we investigate previously unidentified experimental XRD data from the International Centre for Diffraction Data (ICDD) database. The edition we use omits the precise atomic coordinates for all entries and does not clearly specify which structures the coordinates are known for. Hence, our selection of relevant XRD to investigate has used the patterns marked as unidentified in the work presented by Griesemer \textit{et al.} \cite{Griesemer}. We detail four cases here (summarized in Table \ref{table:2}), and results for additional ICDD XRD patterns are provided in the supplementary materials, Ref.~\onlinecite{supplementary}.

$\mathrm{LiSbFe}_2\mathrm{O}_6$ (ICDD ID: 00-044-1073): the ICDD database provides the diffraction pattern, unit cell parameters, and space-group Pnn2, along with the reduced formula unit. The spacegroup 34 has only three Wyckoff positions. Upon running the workflow, we identified that there were 128 
unique Wyckoff protostructures that the given composition can accommodate. We identified that there were two formula units per unit cell, and the protostructure with ID \texttt{A2BC6D\_oP20\_34\_2b\_a\_3c\_a:Fe-Li-O-Sb} was determined to be the solution. This poststructure has 13 degrees of freedom, and our workflow yielded a crystal structure with an \textit{R}-value of 0.09 and an energy of 9 meV/atom above the convex hull of thermodynamic stability from the Materials Project. The entire enumeration workflow was completed in 23 minutes.

A unique Wyckoff protostructure ID does not absolutely ensure that there is no other way to reach the same structure through the already known prototypes, for example, via sub and supergroup relationships and by realizing the crystal structure by placing atoms within the degrees of freedom just slightly on- or off-symmetry in a different prototype. Hence, we calculated formation energies with DFT of all the relevant substitutions into alternative candidate Wyckoff prototypes from both the materials project and WBM to ensure that no other structure reproduced a similar or lower formation energy.
We conclude that the found prototype is a new prototype that is not present in those databases. However, we find a very similar prototype \texttt{ABC2D6\_oP20\_34\_a\_b\_ab\_3c} (Materials Project entry: mp-8673), with a comparable \textit{R}-value of 0.102 which from DFT calculations comes out as slightly less stable at 36 meV/atom above the convex hull. A deeper look into the atomic placements reveals a layered nature of these structures, and it appears the two prototypes represent different stackings between Sb and Fe atoms. (These differences are elaborated on in the supplementary materials, Ref.~\onlinecite{supplementary}, including a radial distribution plot.)

{$\mathrm{SnTe}_2\mathrm{Ni}_3$} (ICDD ID: 00-060-0238): The ICDD entry specifies spacegroup 194 (P63/mmc). Utilizing the \texttt{AflowStringGenerator} method, we identified a total of 74 unique Wyckoff prototypes. The protostructure with ID \texttt{A3BC2\_hP12\_194\_ae\_c\_f:Ni-Sn-Te} was determined to be the solution.
The calculated crystal structure based on this protostructure is 7 meV/atom above the convex hull of thermodynamical stability from the Materials Project, with an \textit{R}-value of 0.059. None of the prototypes present in these datasets met either the R-value or the formation energy criteria. The entire computational workflow was completed in a span of 17 minutes. 

There are no entries with the same protostructure ID \texttt{AB2C3\_hP12\_194\_c\_f\_ae} in the Materials Project or WBM dataset (but we did not do the same exhaustive analysis via DFT calculations of all possible alternatives as for $\mathrm{LiSbFe}_2\mathrm{O}_6$).

{$\mathrm{In}_2\mathrm{Se}_3$} (ICDD ID: 00-040-1408): the pattern files were provided and were identified to be in spacegroup P63/mmc. Using our algorithm, we identify that there are two formula units per unit cell, and the protostructure with ID \texttt{A2B3\_hP10\_194\_f\_af:In-Se} was determined to be the solution, with 2 degrees of freedom along the z direction at the Wyckoff position f, which was occupied by In (f) and Se (a,f). There were only 16 different combinations of Wyckoff positions that satisfied the composition. The \textit{R}-value is 0.23, with an energy above the hull of 55 meV/atom. The corresponding Wyckoff prototype (\texttt{A2B3\_hP10\_194\_f\_af}) is previously reported, as seen in mp-1094034 and mp-1246153.

 $\mathrm{TiFeCl}_3$ (ICDD ID: 00-024-1293). The structure was reported to be in spacegroup 186 (P63mc). Upon enumeration, there were a total of 1874 possible Wyckoff protostructures, from which we identified \texttt{A3BC\_hP10\_186\_c\_a\_b:Cl-Fe-Tl} as the correct protostructure label for the structure with an \textit{R}-value of 0.065, and an energy of 5 meV/atom below the convex the hull. The prototype corresponding to this structure turns out to already exist in the Materials Project (mp-1096852, mp-19241, etc.) and OQMD. Hence, it was also identified by Griesemer \textit{et al.} \cite{Griesemer} with a formation energy insignificantly higher ($3\ \mathrm{meV}/\mathrm{atom}$) than our structure. (There is a slight complication here in that while they also report their match to be in spacegroup P63mc, their final atomic coordinates differ slightly from ours in a way that makes us identify their structure to be in a supergroup, spacegroup 194 (P63/mmc). Hence, it is possible that the structure actually is in this supergroup rather than the spacegroup reported in ICDD.) A comparison of simulated and experimental patterns for these identifcations is shown in Fig. \ref{fig:experimental_results}.

\section{Discussion}

The ability in tests on synthetic patterns to identify prototypes in the structural space omitted in training, along with the identification from previously unidentified experimental XRD patterns from ICDD of two crystal structures with Wyckoff prototypes not present in existing databases, substantiates the capability of our method to identify new, previously unseen, prototypes. The success in applying the method to experimental patterns also demonstrates robustness against noise and other imperfections absent in simulated patterns. Therefore, our approach shows promise not only in theoretical simulations but also in its applicability to real-world experimental data. 

Our test on XRD patterns from ICDD have been successful in finding new structures with better $R$ value matches, and higher stability based on DFT formation energy calculations, compared to solutions using known structures and prototypes. Since the DFT calculations place them below, or just slightly above, the convex hull of thermodynamic stability, well within the theoretical accuracy of DFT with semi-local exchange-correlation functionals of approximately $100\ \mathrm{meV}/\mathrm{atom}$ \cite{kirklin_open_2015, aykolThermodynamicLimitSynthesis2018a} they represent new relevant crystal structures that belong in the materials databases. However, given the non-uniqueness of XRD inversion, further in-depth experimental and theoretical characterization would formally be required to completely confirm these as the true solutions for the respective ICDD entries and dismiss the possibility of even better matches (cf.\ the discussion on limitations in Sec.\ \ref{sec:limitations}).

The identified prototypes could arguably have been found by a type of extended screening suggested in some prior works. For example, the FPASS method by Ward \textit{et al.} \cite{wardAutomatedCrystalStructure2017} describes how to stochastically pivot into a different combination of Wyckoff sites within the same spacegroup during its evolutionary exploration.
However, the difference with the workflow presented in this paper is that it describes how to \emph{systematically} explore the entire space of these previously unseen prototypes.

In Sec.~\ref{sec:limitations}, we have extensively discussed various limitations and challenges in applying our algorithm, including its combinatorial complexity. Nevertheless, our tests demonstrate that our implementation in practice has a computational efficiency sufficient to explore relevant levels of structural complexity. 

\section{Conclusions}
In this work, we have presented a scheme that can resolve experimental XRD patterns without relying on a database of previously resolved structural prototypes.
The use of a coarse-grained descriptor in the Wren ML model allows exploring candidate structures at a low computational effort, which enables prioritizing among large sets of candidates, including in the unexplored parts of the structural space.

We have demonstrated the ability of our model to invert both simulated and experimental XRD patterns.
Our tests show how crystal structures with new Wyckoff prototypes can be obtained, i.e., that symmetric arrangements of atoms that have never been observed before can be identified.
We are unaware of other equally automated and systematic techniques with this capability.
Our work represents a significant advancement in the field, offering a highly automated, efficient, and versatile tool for the identification of new crystal structures. 

\section{Code availability}
The source code is being prepared for public release and is available on request. It will be available as a standalone package and bundled with a future release of \emph{httk}. 

\section{Acknowledgements}
The authors acknowledge insightful discussions with Alpha A. Lee in the early stages of this project, as it spawned out of the development of the Wren machine learning model. A.S.P. and R.A. acknowledge useful discussions related to the work with Florian Trybel. We acknowledge support from the Swedish Research Council (VR) grant no. 2020-05402 and the Swedish e-Science Centre (SeRC). The computations were enabled by resources provided by the Swedish National Infrastructure for Computing (SNIC) at NSC and the National Academic Infrastructure for Supercomputing in Sweden (NAISS) at C3SE, partially funded by the Swedish Research Council through grant agreement no. 2018-05973. Development of the project and calculations were conducted using resource allocations in Berzelius (Berzelius-2022-192) and Alvis (NAISS 2023/22-559).

Author contributions: A.S.P.: Methodology, software, investigation, visualization, and writing the first draft. R.E.A.G.: Conceptualization, Methodology. F.A.F.: Conceptualization, supervision, and writing—review and editing. R.A.: Conceptualization, supervision, and writing—review and editing.

\section{Declaration of conflicts of interest}
FAF is employed by AstraZeneca at the time of publication; however, none of the work presented in this manuscript was conducted at or influenced by this affiliation.

\newpage
\bibliographystyle{apsrev4-2}

\bibliography{main.bib}

\end{document}


\title{Supplementary Materials for: Identifying Crystal Structures from XRD Data using Enumeration Beyond Known Prototypes}

\author{Abhijith S. Parackal}
    \affiliation{Department of Physics, Chemistry and Biology, Linköping University, Sweden}

\author{Rhys E. A. Goodall}
    \affiliation{Department of Physics, University of Cambridge, United Kingdom}

\author{Felix A. Faber}
    \email[Correspondence email address: ]{ff350@cam.ac.uk}
    \affiliation{Department of Physics, University of Cambridge, United Kingdom}
    
\author{Rickard Armiento}
    \email[Correspondence email address: ]{rickard.armiento@liu.se}
    \affiliation{Department of Physics, Chemistry and Biology, Linköping University, Sweden}

\date{\today}

\maketitle

\section{Results}
\subsection{Resolving synthetic XRD to demonstrate generalizability}
In addition to the two cases we investigated, 25 calculations are done from a random sample of structures from the prototype $AB_2C_6$. We note that the below results include rare-earth species for which we know that semi-local DFT calculations are known not to be very accurate, and hence, both the predicted and DFT calculated energies for those materials are likely to have large errors. The columns in the table describe,
\begin{itemize}

\setlength\itemsep{0.005em}
\setlength{\itemindent}{3.0em}
    \item[\textbf{MP-id}]: Materials Project ID of the structure. 
    \item[\textbf{composition}]: alphabetical formula of the structure.
    \item[ \textbf{spgp}]: obtained using \texttt{SpacegroupAnalyser} provided by Pymatgen.
    \item[\textbf{total-enum}]: total number of permitted enumerations, i.e., number of Wyckoff protostructures.
    \item[\textbf{cutoff}]: remaining entries after applying the 40 meV/atom cutoff from the lowest prediction.
    \item[\textbf{true-index}]: the index of the Wyckoff prototype of the original structure (starting from 0).
    \item[\textbf{min-pred}]: lowest energy predicted by Wren for the given list of enumerations (eV/atom)
    \item[\textbf{true-pred}]: predicted energy of the original structure (eV/atom)
    \item[\textbf{max-pred}]: highest energy predicted by Wren for the given list of enumerations (eV/atom)
\end{itemize}

As noted in the workflow session, the 40 meV/atom cutoff is rather strict.
We can see the sensitivity of the cutoff criteria by looking at the spread of predicted energies in the \textbf{max-pred} and \textbf{min-pred} columns.
Users are expected to decide on the cutoff criteria depending on computational resources.
For this example, we can see that some structures (e.g. mp-1207345, mp-765413) would not be resolved under the cutoff criteria (where \textbf{true-index} > \textbf{cutoff}). In summary, 6/28 structures exceed the cutoff criteria, resulting in 78\% recovery in the $AB_{2}C_{6}$-composition-removed Wren model. Curiously, 4 out of the 6 missing compositions belong to the spacegroup 136.

\begin{table}[htbp]
\makebox[\textwidth][c]{
\begin{tabular}{ |p{2cm}||p{3cm}|p{1cm}|p{1cm}|p{1.5cm}|p{1.5cm}|p{1.5cm}|p{1.5cm}| p{1.5cm}|} 
 \hline
     MP-id & composition & spgp & total-enum & cutoff & true-index &  min-pred & true-pred & max-pred \\ \hline
mp-1196973 & Ru4 Tm8 B24 & 55 & 17606 & 8006 & 1495 & -0.658 & -0.642 & 0.703 \\
mp-7791 & Ge2 Li4 F12 & 136 & 106 & 40 & 41 & -2.939 & -2.893 & -2.79 \\
mp-3142 & Ca4 N8 O24 & 205 & 2 & 1 & 0 & -0.883 & -0.883 & -0.660 \\
mp-23047 & Mn1 Bi2 Ho6 & 189 & 47 & 22 & 0 & -0.348 & -0.348 & -0.144 \\
mp-27588 & Be4 P8 O24 & 14 & 21 & 7 & 0 & -2.662 & -2.662 & -2.450 \\
mp-571453 & Sn2 K4 Cl12 & 14 & 7 & 7 & 0 & -1.742 & -1.742 & -1.713 \\
mp-1212905 & Rh4 Er8 B24 & 55 & 17606 & 6151 & 1684 & -0.665 & -0.643 & 0.500 \\
mp-7979 & Pd1 K2 F6 & 164 & 24 & 4 & 4 & -2.344 & -2.297 & -2.168 \\
mp-555747 & W4 Lu8 O24 & 13 & 14904 & 1782 & 758 & -3.276 & -3.245 & -3.109 \\
mp-765413 & Ag2 Li4 F12 & 136 & 106 & 5 & 44 & -2.044 & -1.958 & -1.833 \\
mp-1219800 & P1 Ge2 Ni6 & 189 & 47 & 6 & 10 & -0.319 & -0.238 & -0.107 \\
mp-1200538 & Os4 Tb8 B24 & 55 & 17606 & 10028 & 3932 & -0.612 & -0.591 & 0.757 \\
mp-1206935 & Fe1 Bi2 Tm6 & 189 & 47 & 16 & 1 & -0.376 & -0.375 & -0.188 \\
mp-642315 & Cu4 Te8 Lu24 & 62 & 49 & 29 & 14 & -0.595 & -0.572 & -0.512 \\
mp-1207345 & Pr1 Cs2 F6 & 189 & 47 & 9 & 22 & -3.527 & -3.290 & -3.176 \\
mp-4555 & Ca1 As2 O6 & 162 & 41 & 6 & 0 & -1.785 & -1.785 & -1.272 \\
mp-22097 & Mo2 Rh4 O12 & 136 & 106 & 15 & 23 & -1.348 & -1.290 & -0.930 \\
mp-1205843 & Mn1 Sb2 Er6 & 189 & 47 & 21 & 0 & -0.454 & -0.454 & -0.182 \\
mp-770406 & W4 Sm8 O24 & 19 & 1 & 1 & 0 & -3.135 & -3.135 & -3.135 \\
mp-18191 & Te3 Tl6 O18 & 150 & 7150 & 33 & 18 & -1.132 & -1.110 & -0.524 \\
mp-505545 & Fe4 Lu8 B24 & 55 & 17606 & 6915 & 1897 & -0.664 & -0.645 & 0.565 \\
mp-3188 & Zn2 Sb4 O12 & 136 & 106 & 18 & 38 & -1.678 & -1.598 & -1.321 \\
mp-772617 & Zn4 N8 O24 & 205 & 2 & 1 & 0 & 0.003 & 0.003 & 0.081 \\
mp-976064 & Co1 Te2 Ho6 & 189 & 47 & 9 & 6 & -0.634 & -0.604 & -0.429 \\
mp-610738 & Zr2 In4 Br12 & 128 & 29 & 4 & 3 & -0.906 & -0.884 & -0.655 \\
mp-11755 & Mo1 Sb2 Lu6 & 189 & 47 & 24 & 6 & -0.401 & -0.382 & -0.225 \\
  \hline
\end{tabular}
}
\end{table}

\newpage

\subsection{Recoverability of Wren model on experimental crystal}

Data from the RRUFF database contained structures with partial occupancies and a few structures where simulated XRD  did not match the experimental XRD data. Table \ref{table:cleaning} summarises the filtering applied. 

\begin{table}[ht]
\centering
\begin{tabular}{|l|r|}
\hline
\textbf{Critera} & \textbf{remaining structures} \\
\hline
Number of entries in RRUFF dataset & 2902 \\
After removing partial occupancies & 910 \\
Dropped after spacegroup parsing and filled cell & 539 \\
Less than 64 atoms in a unit cell & 396 \\
R-value less than 0.15 & 320 \\
Less than 16 wyckoff letters & 300 \\
Structures remaining after DFT evaluation & 251 \\
Energy above MP convex hull < 100 meV/atom & 191 \\ 
Completed runs & 189 \\
\hline
Structures correctly identified by our algorithm & 186 \\
\hline
\end{tabular}
\caption{cleaning of RRUFF dataset}
\label{table:cleaning}
\end{table}

\begin{figure*}
    \centering
   \includegraphics[width=0.5\linewidth]{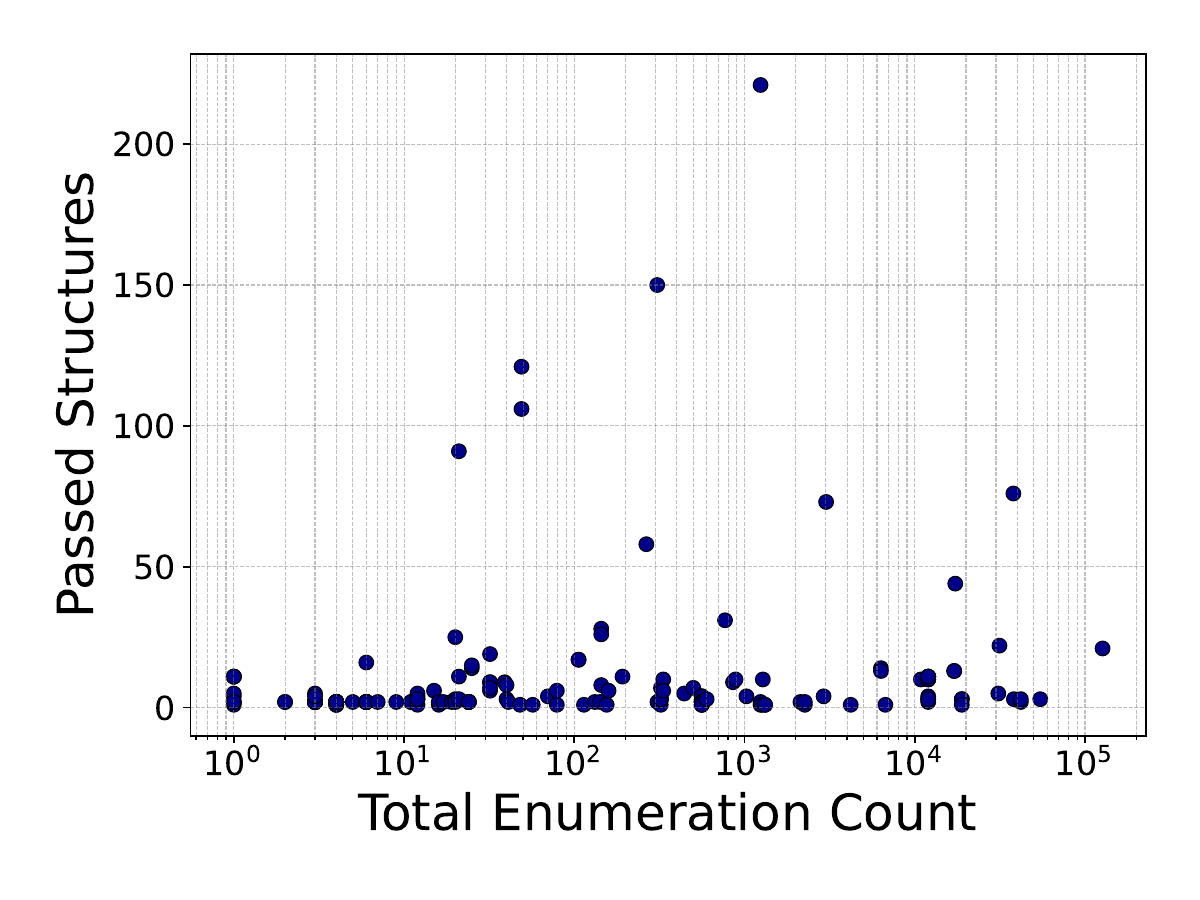}
   \caption{Figure illustrating the effectiveness of the Wren model in identifying experimental crystal structures within the cleaned RRUFF dataset, which comprises 189 experimental XRD data.
   The y-axis represents the total number of enumerations for each composition corresponding to a given XRD, while the x-axis shows the number of structures for each XRD that our workflow deemed to be worthwhile exploring (note that this includes multiple structures per Protostructure, or structures belonging to different Protostructures). This would amount to the number of DFT calculations that must be performed for a given XRD to recover the lowest energy structure. In this particular example, we see that at most 239 crystal structures were evaluated to be viable candidates for a given XRD, with a median of two structures per XRD.   
}
   \label{fig:rruff}
\end{figure*}

Some of the filtering criteria are Wren-specific, such as removing structures with more than 64 atoms in a unit cell and 16 wyckoff sites. We also note that 49 structures could not be calculated using our automated workflow, and 60 structures had a higher energy (100 meV/atom) above the Materials Project convex hull. 

\begin{figure*}
    \includegraphics[width=0.6\linewidth]{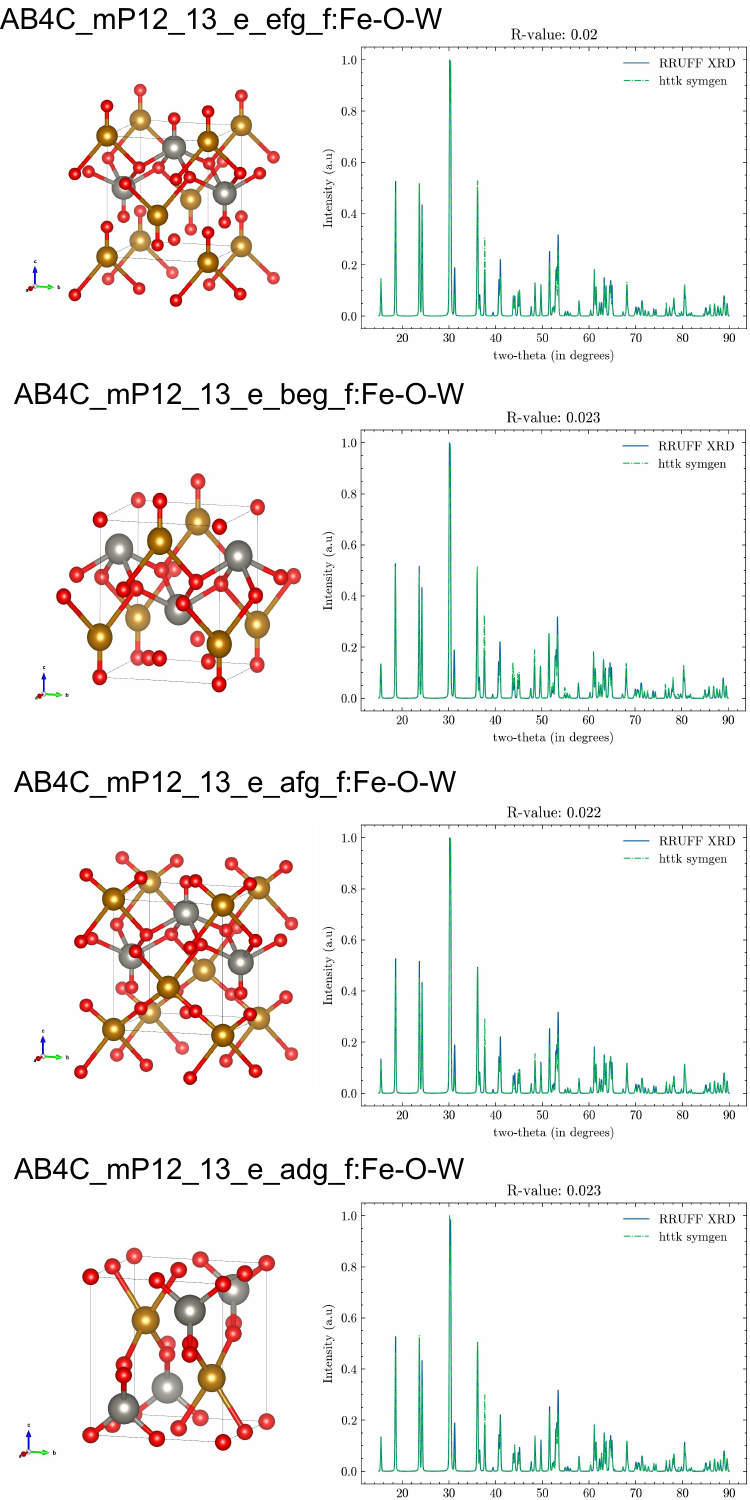}
    \caption{For the experimental XRD data titled \texttt{Ferberite\_R050090-1\_3234}, the workflow identified additional protostructures ( which generated four unique structures) that met the R-value criteria imposed. The simulated XRD pattern aligns well with the XRD pattern provided in the RRUFF dataset, yet the corresponding structures differ drastically in their atomic positions. Identifying the correct crystal structure requires further DFT calculations.}
    \label{fig:duplicate_rruff}
\end{figure*}

Figure \ref{fig:rruff} illustrates the potential speedup achieved by our workflow. When applying our workflow to the RRUFF dataset, the median number of candidate structures per experimental XRD is 2. This represents the number of DFT validations required to determine the correct crystal structure accurately. It is important to note that this number significantly depends on several factors: the XRD cost function cutoff, spacegroup, composition, and the effectiveness of the methods used to identify duplicate structures. Figure \ref{fig:duplicate_rruff} provides an illustration of various crystal structures, all of which correspond to a given experimental XRD.

Upon analysing the three structures that the workflow couldn't identify \ref{table:unidentified structures}, we can see that two structures have four elements in the composition. As discussed above, Wren has trained on the materials project-like dataset, and the distribution of multispecies structures tails off quickly after 3 species in a structure. Nevertheless, this highlights the importance of identifying and incorporating more varied structures into crystal structure databases so that models built on top of them could generalize better.

\begin{table}[htbp]
\makebox[\textwidth][c]{
\begin{tabular}{ |p{9cm}||p{3cm}|p{2cm}|p{2cm}|} 
 \hline
protostructure-ID & total\_enum &  cutoff &   true\_index \\
 \hline
A4BC3D12\_hR60\_155\_ad\_b\_e\_2df:C-Ca-Mg-O        &                114 &             4 &     25 \\
A2B4C\_oI56\_74\_aeg\_2ehj\_h:Mg-O-Si               &              41168 &           258 &   2569 \\
A6B5C3D15\_hP58\_176\_i\_fh\_h\_h2i:Al-B-F-O        &             198723 &          8230 &  19484 \\
  \hline
\end{tabular}
}
\label{table:unidentified structures}

\caption{Stuctures that could not be resolved using Wren screening. \textit{total\_enum} indicates the total number of enumerated prototypes for the composition in the given spacegroup; cutoff indicates the last index that remains after sorting the enumerations using wren and applying the 40 meV/atom cutoff from the lowest prediction. \textit{true\_index} is the position of the true protostructure (the ID given in the protostructure-ID column) after Wren screening, and it is higher than the cutoff index implies we would not be considering the structure for our workflow.}
\end{table}

\newpage

\subsection{Structure solutions of experimental XRD from ICDD database}

Table of supplementary results for the tests presented in the paper for resolving experimental XRD. We note that the results below include rare-earth species for which we know that semi-local DFT calculations are known not to be very accurate, and hence, both the predicted and DFT calculated energies for those materials are likely to have large errors. The columns are

\begin{itemize}

\setlength\itemsep{0.005em}
\setlength{\itemindent}{9.0em}
    \item[\textbf{ICDD/id}]: The ID in the ICDD of XRD. 
    \item[\textbf{protostructure ID}]: Our ID to identify prototypes with assigned species (see paper).
    \item[ \textbf{DFT hull}]: the positive or negative distance to the known convex hull of thermodynamic stability from Materials Project; i.e., when negative, it indicates a structure that is stable with respect to the competing phases, and thus that the know hull needs to be updated with the listed structure.
    \item[\textbf{R-value}]: the $R$ measure of how well the predicted structure matches the experimental XRD.
\end{itemize}

\begin{table}[htbp]
\makebox[\textwidth][c]{
\begin{tabular}{ |p{3cm}||p{8cm}|p{2cm}|p{2cm}|} 
 \hline
 ICDD-ID & protostructure ID & DFT hull (meV/atom) & R-Value  \\ \hline
000501046 & ABC2\_hR12\_166\_a\_b\_c:Ho-K-Te & -121 & 0.142 \\
000341392 & A3BC6\_cF40\_225\_ac\_b\_e:Cs-Dy-F & -1 & 0.084 \\
000300467 & AB4C\_oP24\_19\_a\_4a\_a:Al-Cl-Cu & 43 & 0.334 \\
000240913 & AB6C2\_tI18\_121\_a\_di\_e:Mo-O-Pr & 64 & 0.206 \\
000600235 & AB2C\_oP16\_62\_c\_2c\_c:Al-Ti-Zr & 66 & 0.217 \\
000381294 & A6B2C\_hP9\_162\_k\_c\_b:F-Li-Pr & 72 & 0.168 \\
000420410 & AB3C\_hR30\_148\_c\_f\_ab:Cr-O-Rh & 86 & 0.098 \\
000391131 & A5B2C\_tI32\_140\_cl\_h\_a:I-Pb-Tl & 107 & 0.136 \\
000510892 & A4B2C5\_tI22\_107\_2ab\_2a\_a2b:Cu-Pr-Sn & 140 & 0.241 \\
000211080 & ABC2\_oP16\_33\_a\_a\_2a:Ag-Fe-O & 158 & 0.079 \\
  \hline
\end{tabular}
}
\caption{Calculated \textit{R}-value and Materials Project hull distance for a list of unidentified XRD patterns in the ICDD 2012 database.}
\end{table}

\newpage

\section{Analysis of {$\mathrm{LiSbFe}_2\mathrm{O}_6$}}

\begin{figure*}
    \centering
    \includegraphics[width=0.6\linewidth]{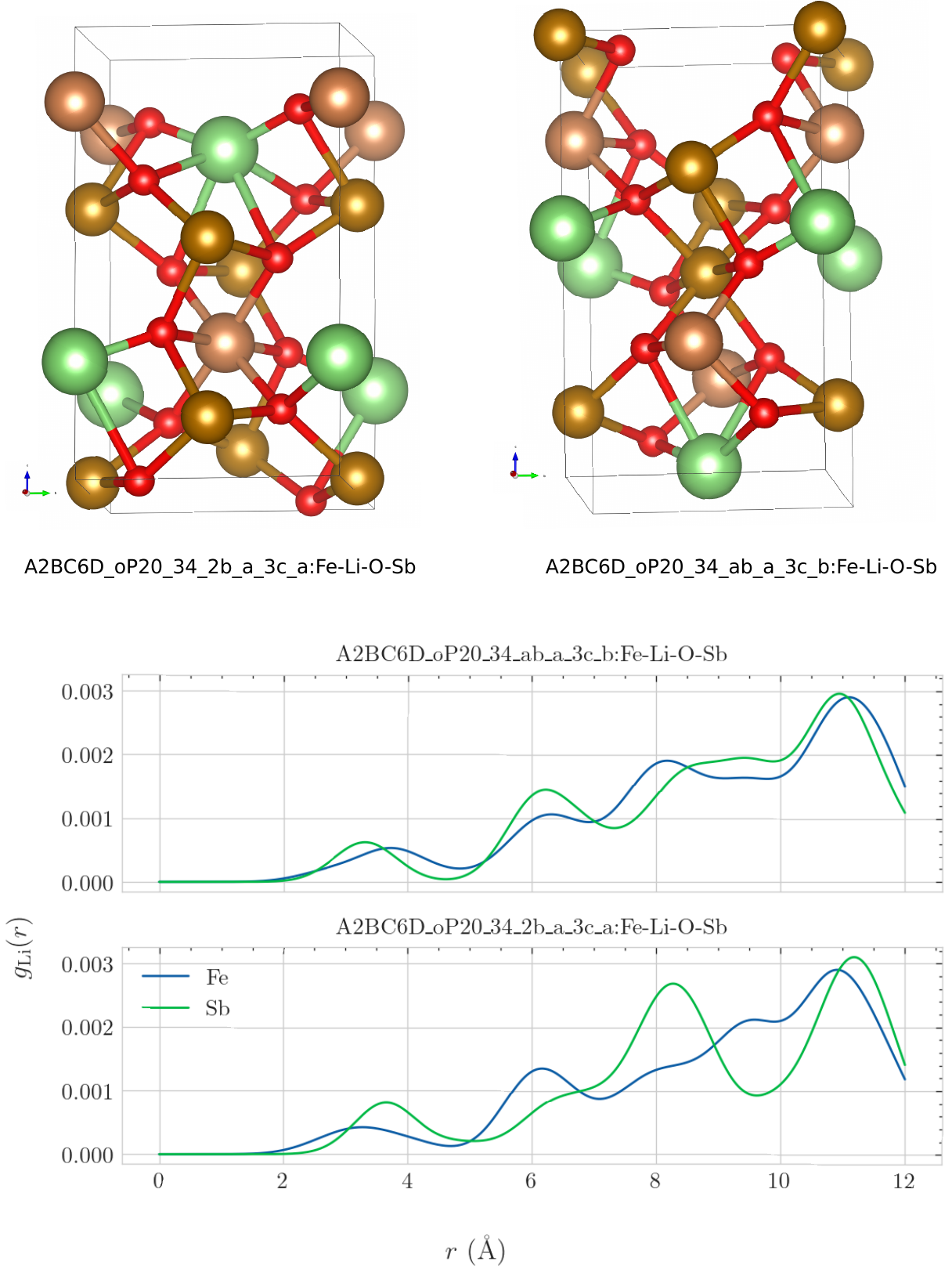}
    \caption{Comparison of Two Wyckoff Prototypes: The structure on the left represents a Wyckoff prototype not found in any existing database. The two prototypes differ only in two specific Wyckoff positions. Despite this similarity, the local environments surrounding the Lithium atoms in each structure are distinct. To analyze this, we identify the nearest neighbours of a given Lithium atom within a radius of 12 Ångströms. A Gaussian function is applied to each point, and the results are aggregated to produce a Radial Distribution Function (RDF) that characterizes the neighbouring environments. In this figure, the RDFs for Lithium's interactions with Iron and Antimony are plotted. The differences in the RDF profiles highlight the variations in the local environments between the two structures.
 }
    \label{fig:rbf-plot}
\end{figure*}

The protostructure \texttt{A2BC6D\_oP20\_34\_2b\_a\_3c\_a:Fe-Li-O-Sb} was found to be the appropriate candidate for the composition {$\mathrm{LiSbFe}_2\mathrm{O}_6$}, and this Wyckoff prototype does not exist in the Materials Project + WBM dataset. Upon investigation, an existing prototype \texttt{ABC2D6\_oP20\_34\_a\_b\_ab\_3c} yields a protostructure \\ \texttt{A2BC6D\_oP20\_34\_ab\_a\_3c\_b:Fe-Li-O-Sb}, which gives a crystal structure with a similar  \textit{R}-value (0.1) and is 30 meV/atom above the convex hull. These two prototypes only differ in the Wyckoff positions of Fe and Sb, where one of the Fe occupies an 'a' site, while Sb occupies a 'b' site.

The second wyckoff prototype yields a structure that is deceptively similar to the one that we obtained. However, we can see that a radial distribution function of the interatomic separation of Lithium (Fig: \ref{fig:rbf-plot}) against neighbouring Iron and Antimony atoms gives different profiles, implying that the neighbouring environment of Lithium atoms in both of these structures is qualitatively different.

\subsection{Enumeration of Wyckoff positions}
Below are the algorithmic representations of the Python code used to generate Wyckoff protostructures.
This is broken down into two routines, one which solves the non-negative integer solution for a linear diophantine equation and another which calls on Alg \ref{alg:linerdop} to obtain said solutions and convert them into a sequence of Wyckoff letters.
Checking and removing duplicated Wyckoff sites with no degrees of freedom are done in subsequent steps (but are not given here). In the implementation, both algorithms are written using Python generator functions.

For Alg \ref{alg:recursive_solution}, for a given spacegroup, we first group the Wyckoff letters by whether it has at least degrees of freedom and then by their multiplicity.
When Alg \ref{alg:linerdop} is called, it is only expected to solve the unknowns for a small number of coefficients. Therefore, the brute force method implemented in Alg \ref{alg:linerdop} does not contribute to the combinatorial wall one would face for larger compositions.
The contributing factor for the combinatorial impasse is caused by the \texttt{combinations without replacements} and \texttt{combinations with replacements} and is very notable for spacegroups with a large number of Wyckoff sites, like spacegroup 47 (with 27 Wyckoff sites).

\begin{algorithm}[H]
\begin{algorithmic}
\caption{Enumerate all Wyckoff assignments for one element. In practice, the nested loops can be implemented via recursion or as a product of iterators (see source code).}\label{alg:recursive_solution}
\Input
\State \emph{sitegroups}: List of list of letters: Wyckoff label groups that group labels with the same multiplicity.
\State \emph{multiplicities}: List of integers: multiplicity for each label group.
\State \emph{dofs}: List of bool: \emph{true} for site groups for Wyckoff sites with at least one degree of freedom, otherwise \emph{false}.
\State \emph{count}: The number of atoms of the specific element to assign to Wyckoff sites
\EndInput
\Output
\State \emph{siteassignments}: List of List of (Integer, Letter): number of atoms occupying specific Wyckoff sites.
\EndOutput
        \State \emph{N} $\gets$ len(\emph{sitegroups})    
        \State \emph{siteassignments} $\gets []$    
        \State \emph{maxvals} $\gets []$    
        \For{$i$ \textbf{in} 0...len(\emph{dofs})}
            \If{\emph{dofs}[$i$]}
              \State \emph{maxvals}[$i$] $\gets$ $\infty$
            \Else
              \State \emph{maxvals}[$i$] $\gets$ len(\emph{sitegroups}[$i$])
            \EndIf            
        \EndFor          
        \State \emph{solutions} $\gets$ all Diophantine eqn.\ solutions for \emph{coeffs} = \emph{multiplicities}, \emph{target} = \emph{count}, \emph{maxvals} \Comment{See Algorithm 2}
        \For{$solution$ \textbf{in} \emph{solutions}}
            \If{$\textit{dofs}$[0]}
              \State \emph{assignments}[0] $\gets$ all combinations without replacements of \emph{sitegroups}[0] and \emph{solution}[0]
            \Else
              \State \emph{assignments}[0] $\gets$ all combinations with replacements of \emph{sitegroups}[0] and \emph{solution}[0]
            \EndIf            
            \For{\emph{assignment}[0] \textbf{in} \emph{assignments}[0]}
                \If{\emph{dofs}[1]}
                   \State \emph{assignments}[1] $\gets$ all combinations without replacements of \emph{sitegroups}[1] and \emph{solution}[1]
                \Else
                   \State \emph{assignments}[1] $\gets$ all combinations with replacements of \emph{sitegroups}[1] and \emph{solution}[1]
                \EndIf
                \For{\emph{assignment}[1] \textbf{in} \emph{assignments}[1]}
                  \State ...
                    \For{\emph{assignment}[$N-1$] \textbf{in} \emph{assignments}[$N-1$]}
                       \State append [\emph{assignment}[0], \emph{assignment}[1], ... \emph{assignment}[$N-1$]] to \emph{siteassignments}
                    \EndFor
                \EndFor
            \EndFor
        \EndFor
\end{algorithmic}
\end{algorithm}

\begin{algorithm}[H]
\begin{algorithmic}
\caption{Enumerate all constrained positive solutions to the linear integer Diophantine equation: find all sets of parameters to multiply with the non-negative integer coefficients in \emph{coeffs} to make the total sum: $ax + by + cz + ... = target$, such that all parameters are whole numbers less than or equal to the corresponding value in \emph{maxvals}. In practice, the nested loops can be implemented via recursion or as a product of iterators (see source code).}
\label{alg:linerdop}
\Input
\State \emph{coeffs}: List of the given coefficients.
\State \emph{target}: The right-hand side target value for the Diophantine equation.
\State \emph{maxvals}: List of maximum parameter values (positive integer or infinite).
\EndInput
\Output
\State \emph{solutions}: List of list of parameters that solve the equation.
\EndOutput
    \State \emph{solutions} $\gets []$    
    \State \emph{N} $\gets$ len(\emph{coeffs})    
    \State \emph{targets}[0] $\gets \textit{target}$    
    \State \emph{maxparams}[0] $\gets \min(\lfloor \textit{targets}[0]/\textit{coeffs}[0] \rfloor, \textit{maxvals}[0])$    
    \For{\emph{params}[0] \textbf{in} 0...\emph{maxparams}[0]}
        \State \emph{targets}[1] $\gets \textit{targets}[0] - \textit{coeffs}[0]\cdot\textit{params}[0]$           
        \State \emph{maxparams}[1] $\gets min(\lfloor \textit{targets}[1]/\textit{coeffs}[1] \rfloor$, \emph{maxvals}[1])    
        \For{\emph{params}[1] \textbf{in} 0...\emph{maxparams}[1]}
            \State ...
            \For{\emph{params}[$N-1$] \textbf{in} 0...\emph{maxparams}[$N-1$]}
                \If {$\textit{targets}[N-1] - \textit{coeffs}[N-1]\cdot\textit{params}[N-1] = 0$}
                     \State append \emph{params} to \emph{solutions}
                \EndIf
            \EndFor
        \EndFor
    \EndFor
\end{algorithmic}
\end{algorithm}